\documentstyle[aps,preprint,epsfig]{revtex}
\begin{document}
\preprint{                                \vbox{\hbox{CERN-TH/96-216}
                                                \hbox{LBNL-39237}}    }
\draft
\title{       Indications from Precision Electroweak Physics\\
           Confront Theoretical Bounds on the Mass of the Higgs Boson } 
\author{	              John Ellis%
\thanks{ This work was supported in part by the Director, Office of 
	 Energy Research, Office of Basic Energy Science of the U.S. 
	 Department of Energy, under Contract DE-AC03-76SF00098.}     }
\address{          CERN, CH-1211, Geneva, Switzerland                 }
\author{                G.~L.~Fogli and E.~Lisi                       }
\address{ Dipartimento di Fisica and Sezione INFN di Bari, Bari, Italy}
\maketitle
                           \begin{abstract}
An updated fit to the precision electroweak data and to the direct
measurement of the top quark mass $m_t$ provides significant constraints 
on $m_t$ and on the Higgs boson mass $M_H$: $m_t/\text{GeV}=172\pm 6$
and $\log_{10}(M_H/\text{GeV})=2.16\pm 0.33$, with an error correlation
$\rho=0.5$. We integrate the $(M_H,\,m_t)$ probability distribution  
found in this analysis over various zones of the  $(M_H,\,m_t)$ plane
defined by one-sided experimental and theoretical bounds on the Higgs 
boson mass, both in the Standard Model and in its minimal supersymmetric 
extension. The comparison of the cumulative probabilities gives 
interesting information on the likelihood that the true value of
$M_H$ is compatible with different theoretical scenarios.   
\end{abstract}

\newpage

	The possibility of constraining {\it both\/} the top quark mass 
$m_t$ {\it and\/}  the Higgs boson mass $M_H$ through their virtual 
effects on precision electroweak observables was recognized long ago
\cite{El90}. The  continuous refinement of the experimental measurements 
at the CERN Large Electron Positron Collider (LEP) and elsewhere has 
resulted in sustained improvement of these ``indirect'' bounds on $m_t$ and 
$M_H$, both in the Standard Model (SM) \cite{El91,El93,Ag92,No93,Ha94,Mo94}
and in the Minimal Supersymmetric extension of the Standard Model (MSSM) 
\cite{El92,El94,Li93,Ch96,Ho96}. The predictive power of the precision 
electroweak data was confirmed dramatically by the ``direct'' determination 
of the top quark mass by the CDF \cite{Ctop} and D0 \cite{Dtop} experiments 
at the Tevatron, which currently yield $m_t=175\pm 6$ GeV (as reported in 
\cite{EWWG}). Combining these measurements with the precision electroweak 
data enables the corresponding prediction of $M_H$ to be improved 
significantly \cite{El96,Ch95,Ma95}.

	A new stage has recently been attained with the release of
preliminary new data from LEP, including almost all the data taken with 
Phase 1 of LEP around the $Z^0$ peak. The most recent available electroweak 
precision data from LEP and SLC are reported in \cite{EWWG}.  In this paper, 
we first update our previous analyses of the precision electroweak data, 
combining the new data with the older low-energy precision data as described 
in \cite{El96}. We then confront the resulting fit with one-sided 
experimental and theoretical bounds on $M_H$ in the Standard Model and 
the MSSM. The experimental bounds  come from unsuccessful direct searches 
for the  standard or supersymmetric Higgs boson at LEP~1 \cite{PDGR,Gr95}.
Theoretical bounds in the Standard Model come from requiring its validity 
up to some large scale $\Lambda$, below which the current electroweak vacuum 
is assumed to be metastable \cite{Es95}, and renormalization group evolution 
does not cause the Standard Model couplings to diverge \cite{Li86,Sh89}. 
Theoretical bounds within the MSSM come from calculations of the mass of 
the lightest neutral Higgs boson, including quantum corrections \cite{MSSM}.
These bounds divide the $(M_H,\,m_t)$ plane into several regions, depending 
on their consistency or otherwise with the Standard Model and/or the MSSM.

	We use the joint probability distribution of $(M_H,\,m_t)$ provided 
by our new global fit to estimate the relative (cumulative) probabilities 
that the true values of $M_H$ and $m_t$ lie within each of these different 
regions. This enables us to estimate the likelihood that the true values of 
$M_H$ and $m_t$ will turn out to be compatible with the Standard Model and/or
the MSSM. There is at present no significant difference between the 
likelihoods of the Standard Model and the MSSM, but a refinement of this 
type of analysis with future improvements in the precision electroweak 
data set has the potential to provide some discrimination between these
models.

	We start by reporting the result of a global analysis 
within the SM of the precision electroweak data recently made available,
excluding initially the direct $m_t$ determination. Our fitting program 
also includes all available lower-energy precision data, along the lines 
described in \cite{El96}. It now yields
\begin{eqnarray}
m_t/\text{GeV}&=& 157^{+16}_{-12}\ ,\\
\log_{10}(M_H/\text{GeV})&=& 1.81^{+0.45}_{-0.36}\ ,
\end{eqnarray}
where the errors on both $m_t$ and log$_{10}(M_H/$GeV) are at the 1-$\sigma$ 
level. The information on $M_H$ is quoted on a logarithmic scale, because 
electroweak observables characteristically exhibit a logarithmic dependence 
on $M_H$, and the probability distribution we find is closer to being 
Gaussian in log$_{10}(M_H/$GeV). The corresponding numerical values of $M_H$ 
at the 1-$\sigma$ level are $M_H\,=\,65^{+117}_{-37}$ GeV. We recall that
this estimate of $M_H$ was obtained without using the CDF and D0 measurements 
of $m_t$. It is consistent with the indications for a light Higgs mass 
obtained in our previous works \cite{El96,El94}, as well as in 
\cite{Ch96,Ch95}.

	The estimate (1) of $m_t$ is less than 1-$\sigma$ below the direct
measurement by the CDF and D0 collaborations:  $m_t=174 \pm 6$ GeV. This 
agreement constitutes dramatic confirmation of the SM at the one-loop level. 
It also justifies combining \cite{El96,El94} the indirect and direct 
measurements of $m_t$, which has the effect of readjusting the previous 
best-fit range (1) to higher values of $m_t$. The well-known positive
$m_t$-$M_H$ correlation in the radiative corrections then causes the
best-fit value of $M_H$ to increase as well:
\begin{eqnarray}
m_t/\text{GeV}&=&172\pm6\ , \\
\log_{10}(M_H/\text{GeV})&=&2.16\pm0.33\ .
\end{eqnarray}
We observe that the error in $\log(M_H)$ is now somewhat reduced and
more symmetrical, and that the combined probability distribution is 
approximated to a good accuracy by a bivariate Gaussian in the variables
$x=\log_{10}(M_H/\text{GeV})$ and $y=m_t/\text{GeV}$. This best-fit 
Gaussian distribution, that will be used hereafter, is completely defined 
by the 1-$\sigma$ errors in Eqs.~(3) and (4) and  by their correlation,
which in our fit is $\rho_{xy}\,=\,0.5$. The 1-$\sigma$ range (4) 
corresponds to $M_H\,=\,145^{+164}_{-77}$ GeV. The results of our global 
fit are in good agreement with those recently reported by the LEP 
Electroweak Working Group \cite{EWWG}.

	The joint bounds on $(x,\,y)$ in the MSSM do not differ appreciably
from those in the SM, as long as the the MSSM spectrum is sufficiently heavy 
to be decoupled \cite{El94,El96}. We assume the MSSM parameters 
$m_{\tilde{g}}=m_0=-\mu= 1$ TeV in the following (notation as in \cite{Li93}),
so that this decoupling is enforced. Subleading terms in the radiative 
corrections induce small differences between the MSSM and the SM only at 
low values of $M_H$, which are anyway disfavored by the probability 
distribution of $M_H$ itself [Eq.~(4)]. These observations indicate that, 
given the present information, it is reasonable to use the same 
$(x,\,y)$ probability  distribution in the SM and MSSM, subject to the 
different one-sided experimental and theoretical bounds that we discuss now.

	In Fig.~1 we show the 1-$\sigma$ and 2-$\sigma$ contours 
$(\Delta\chi^2=1,\,4)$ of the joint probability distribution in the plane 
$(\log M_H,\,m_t)$ (solid ellipses), together with experimental and 
theoretical one-sided bounds applicable in the Standard Model. The vertical 
hatched line represents the LEP lower bound $M_H > 65$ GeV \cite{PDGR}. 
The sloping curves on the left represent the lower limits on $M_H$ coming 
from the requirement of `metastability' of the  electroweak vacuum 
\cite{Es95}: their slopes  reflect the dependence of this type of bound 
on $m_t$. The different curves correspond to the requirement that our 
present electroweak vacuum have a lifetime exceeding $10^{10}$ years for 
any transition to a lower-lying state with a Higgs expectation value
$|H|\,\le\,\Lambda$, according to calculations with the renormalization 
group improved effective potential. The curves on the right  represent 
the upper limits on $M_H$ derived by Lindner \cite{Li86} from the 
`triviality' requirement that none of the SM couplings should become 
singular at any renormalization scale $\mu\,\le\,\Lambda$. Taken together, 
these two sets of lines represent the requirement that the SM remain 
consistent at all scales below $\Lambda$. We observe that, for any given 
value of $\Lambda$, there is only a relatively narrow vertical band, 
narrowing at high $m_t$, which is allowed in Fig.~1 by the theoretical 
and experimental bounds.

	In order to infer any useful information about the relative 
likelihoods that the true values of $(\log M_H,\,m_t)$ will be consistent 
with different values of $\Lambda$, it is necessary to take into account the
joint $(\log M_H,\,m_t)$ probability distribution, as shown in Fig.~2.

	In Fig.~2 we report the integrated (cumulative) probability in the 
region of the $(\log M_H,\,m_t)$ {\it excluded} by LEP,  by vacuum
metastability, and by triviality. The complementary fraction of probability 
gives the cumulative probability that the true value of $(\log M_H,\,m_t)$ 
lies in the allowed region. This exercise is repeated for different values 
of $\Lambda$. We notice that the LEP bound excludes only 15\% of the total 
probability, providing the non-trivial information that the global fit to 
the precision electroweak data is statistically consistent with the negative 
results of the searches  for the Higgs boson at LEP. On the other hand, the 
metastability  bound excludes a significant fraction of the probability, 
unless $\Lambda \simeq 10^4$ GeV. In this case the metastability bound 
only excludes a zone which is already almost completely forbidden by the LEP
searches (see Fig.~1).  The triviality bounds also exclude a fraction of 
the probability which increases with $\Lambda$. The remaining allowed region 
is  weighted by a cumulative probability which decreases from 77\% at low
$\Lambda$ to 27\% at high $\Lambda \simeq 10^{19}$ GeV.

	Since the area of the $(\log M_H,\,m_t)$ plane that is allowed 
for large $\Lambda$ is included within that allowed for small $\Lambda$, it 
is inevitable that the cumulative probability decrease monotonically with 
$\Lambda$. If the decrease is gradual, no useful information about the 
likelihood of different values of $\Lambda$ can be extracted, whereas a 
precipitous decrease would indicate that some range was highly disfavored.
We see from Fig.~2 that current data do not exclude statistically
any value of $\Lambda$. However, according to the available information, 
it is about three times more likely that the true values of 
$(\log M_H,\,m_t)$ are consistent  with a Standard Model valid up to a 
scale of $10^4$ GeV than up to the Planck scale. This is an interesting 
piece of information that will become more specific as further constraints 
are placed on $(\log M_H,\,m_t)$, culminating in the eventual direct 
measurement of $M_H$.

	In Fig.~3 we study analogous constraints in the MSSM.  In this case, 
the LEP lower limit on $M_H$ varies with the ratio of supersymmetric Higgs 
vacuum expectation values, $\tan\beta$ \cite{Gr95}.  The upper theoretical 
bound on $M_H$ \cite{MSSM} also varies with $\tan\beta$ as well as with 
$m_t$. We do not show in Fig.~3 the lower theoretical bound on $M_H$,
which is always below the LEP bounds.

	In Fig.~4 we show the cumulative probabilities obtained by integrating
the differential $(\log M_H,\,m_t)$ distribution in the various zones of 
Fig.~3. This exercise is repeated for different values of $\tan\beta$,
and we see that a large fraction of the probability is excluded by the upper 
theoretical bound. This is a consequence of the best-fit value of $M_H$,
which is at the limits of the allowed region in Fig.~3. No value of 
$\tan\beta$ can be statistically excluded, and we have not explored the 
quality of fits away from the decoupling limit of large sparticle mass
parameters. However, it currently appears more likely that the true values 
of $(\log M_H,\,m_t)$ are consistent with a MSSM with high $\tan\beta$ 
($\gtrsim 8$) than with $\tan\beta\simeq 1$.

	It has been noticed \cite{Di95,Es96} that the bounds on $M_H$ in 
the SM and MSSM  define some zones in the $(\log M_H,\,m_t)$ plane where 
only one of the models (either the SM or the MSSM) is allowed. In other 
zones the SM and MSSM are both consistent, and the discovery of a Higgs 
boson would not help to discriminate between the models. We have made an 
exploratory calculation of the cumulative probabilities  that the true 
values of $(\log M_H,\,m_t)$ lie in each of these zones, as shown in an 
SM-MSSM ``phase diagram'' in  Fig.~5 for the particular cases 
$\Lambda=10^{19}$ GeV and $\tan\beta=4$.

	In Fig.~5,  the zone labelled 1 is bounded by the LEP lower limit 
on $M_H$ in the MSSM. Zone 2 is bounded by this  limit and by the LEP lower 
limit on $M_H$ in the SM. Zones 3, 4, 5 and 6 are bounded by the strongest 
of the upper or lower SM or MSSM theoretical constraints. Zone 7 is excluded 
by triviality in  the SM. The current probabilities that, according to the 
most complete available information, the true values of $(\log M_H,\,m_t)$ 
lie in each of the various zones are also indicated. Apart from zone 7, 
the two regions that appear most likely are 3 and 6, within which the SM 
and MSSM can be distinguished. Zone 3 corresponds to values of $M_H$ that are
above the LEP limit for the SM Higgs {\it but\/} below the SM vacuum 
metastability bound, {\it though\/} below the MSSM upper bound. Zone 6
corresponds to values of $M_H$ allowed in the SM  but above the upper limit 
imposed in the MSSM.

	The zones in which the SM is consistent are 4 and 6, and we estimate 
a cumulative probability of 27\% that the true values of $(\log M_H,\,m_t)$ 
lie within one or the other of these zones. The zones in which the MSSM is 
consistent are 2, 3, and 4, and we estimate a cumulative probability of 32\% 
that the true values of $(\log M_H,\,m_t)$ lie within one of these zones. 
Note that the likelihood of the MSSM zones is not lower than that of the SM 
zones, even though the central value of $M_H$ lies well inside the region
of Fig.~1 that is consistent with the SM. Clearly, both the SM and the 
MSSM are highly consistent with the present data, which cannot be said 
to favour either of them in a significant way.

	Looking to the future, however, there is the prospect that 
improvements in the precision electroweak data set, in particular greater 
accuracy in the $M_W$ measurement \cite{Ka95},  could provide some useful 
indication one way or the other. Also, any direct  measurement of $M_H$ 
may well resolve the issue. However, this is not guaranteed, since there 
is a region in Fig.~5, namely zone 4, where measured values of $M_H$ and 
$m_t$ would be consistent with both the SM and the MSSM. The cumulative
probability that the true value of $(M_H,\,m_t)$ lie in this zone 
(around $5\%$) is not completely negligible.

In conclusion: we have analysed the most complete available 
information from precision electroweak measurements
to determine the $(M_H,\,m_t)$ probability distribution. 
We have used the best-fit Gaussian approximation to this distribution to
evaluate the cumulative probability that the true values of  $(M_H,\,m_t)$
are consistent with the experimental and theoretical one-sided bounds
on $M_H$, both in the SM and in MSSM. Both the SM and the MSSM are consistent
with the available data and the known constraints.

	J.E.\ thanks Mike Chanowitz and Hitoshi Murayama for useful 
discussions,  and the LBNL Theoretical Physics Group and the Berkeley Center 
for Particle  Astrophysics for kind hospitality: his work was supported in 
part by  the Director, Office of Energy Research, Office of Basic Energy 
Science of the U.S.\ Dep.\ of Energy, under Contract DE-AC03-76SF00098.


\begin{figure}
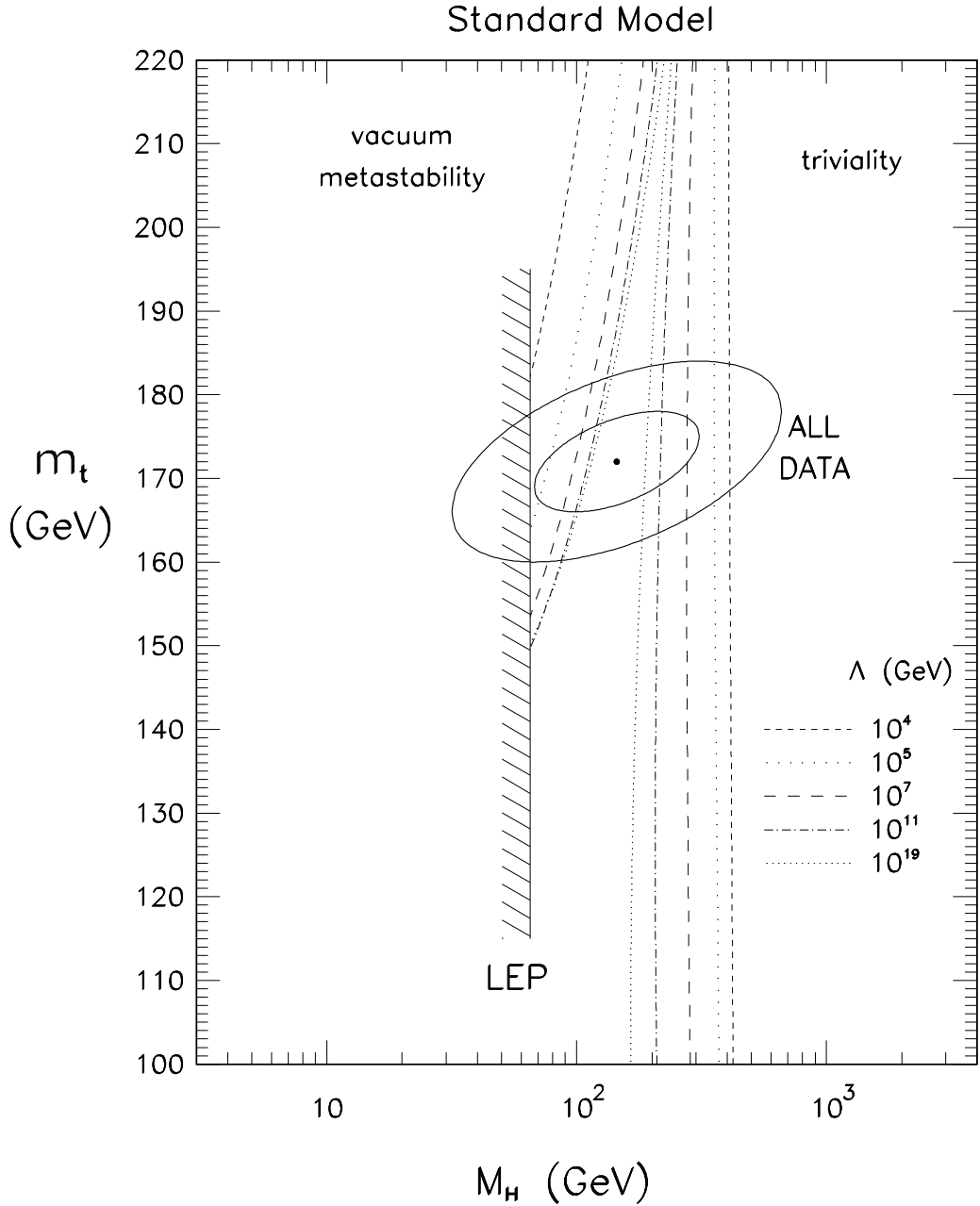

\caption{	Indirect bounds on $(M_H,\,m_t)$ and one-sided experimental
		and theoretical limits in the Standard Model.  The solid 
		ellipses represent the 1-$\sigma$ and 2-$\sigma$ contours 
		from the best-fit  Gaussian distribution obtained by 
		analysing all electroweak precision data, including the 
		measurement of $m_t$ at CDF and D0. The hatched line is
		the LEP lower bound on $M_H$ \protect\cite{PDGR}. 
		The other curves represent the lower and upper limits on 
		$M_H$ from vacuum metastability  \protect\cite{Es95} and 
		triviality \protect\cite{Li86,Sh89} respectively, as 
		functions of the scale of new physics $\Lambda$.}
\end{figure}
\begin{figure}
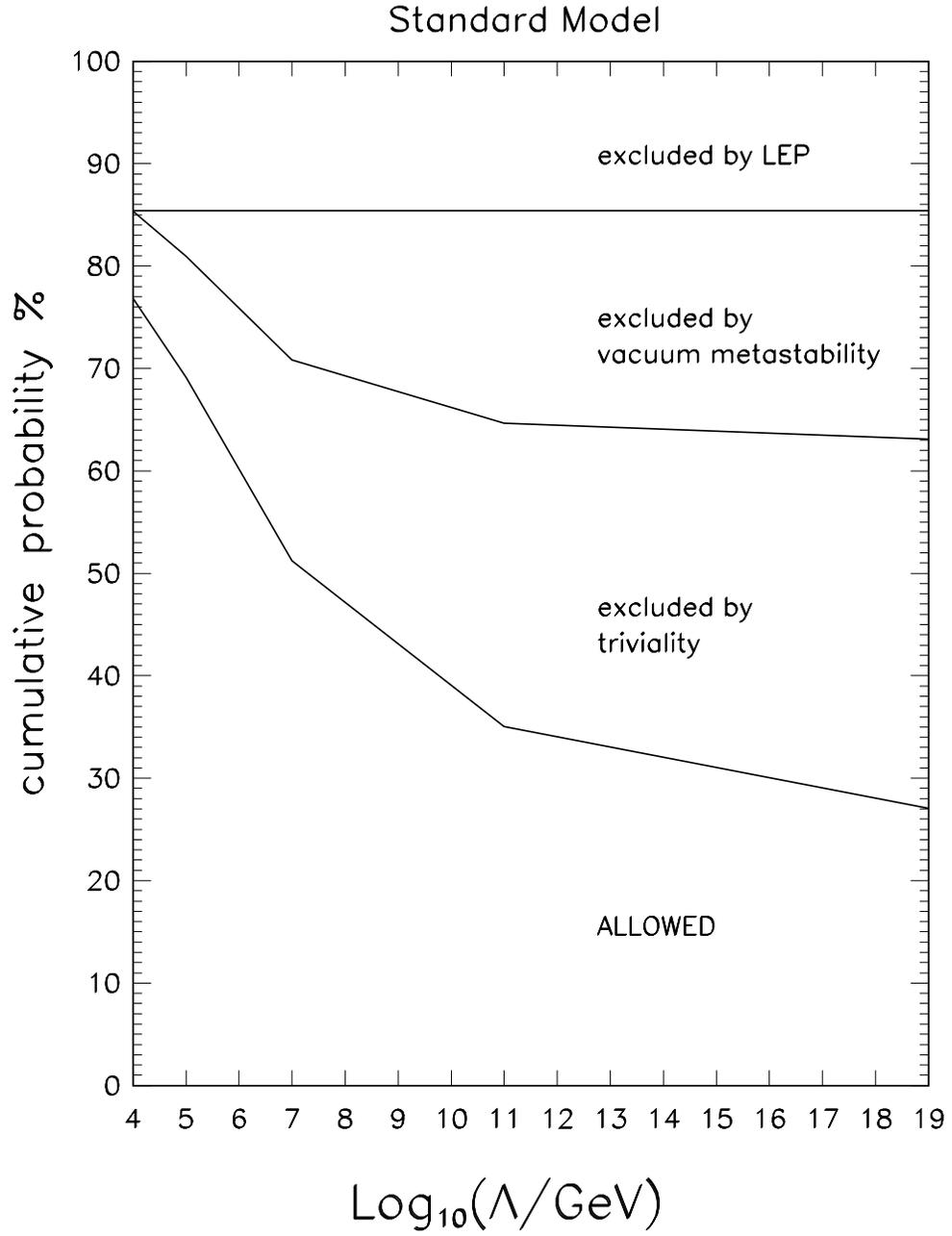

\caption{	Cumulative (integrated) probabilities in the various zones
		excluded or allowed in the Standard Model by one-sided 
		bounds. No value of $\Lambda$ can be excluded.}
\end{figure}
\begin{figure}
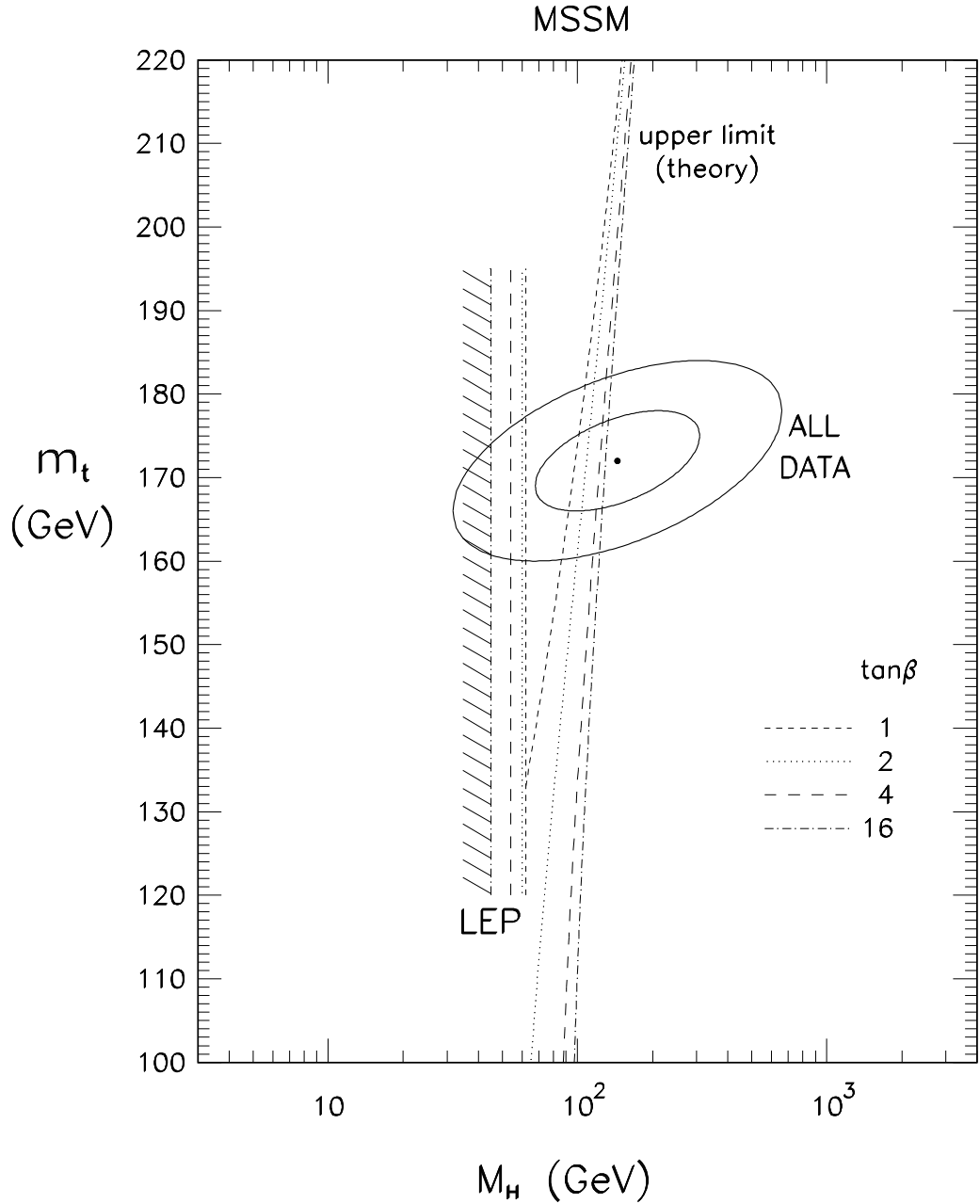

\caption{	Indirect bounds on $(M_H,\,m_t)$ and one-sided experimental
		and theoretical limits in the Minimal Supersymmetric 
		Standard Model.  Apart from the Higgs sector, the MSSM 
		spectrum is assumed to be decoupled. Solid ellipses represent 
		the 1-$\sigma$ and 2-$\sigma$ contours as in Fig.~1. 
		The vertical lines are the LEP lower bounds on $M_H$ 
		\protect\cite{Gr95}, which depend slightly on $\tan\beta$. 
		The other curves represent the upper limits on $M_H$ in 
		the MSSM \protect\cite{MSSM}, as  a function of $\tan\beta$.}
\end{figure}
\begin{figure}
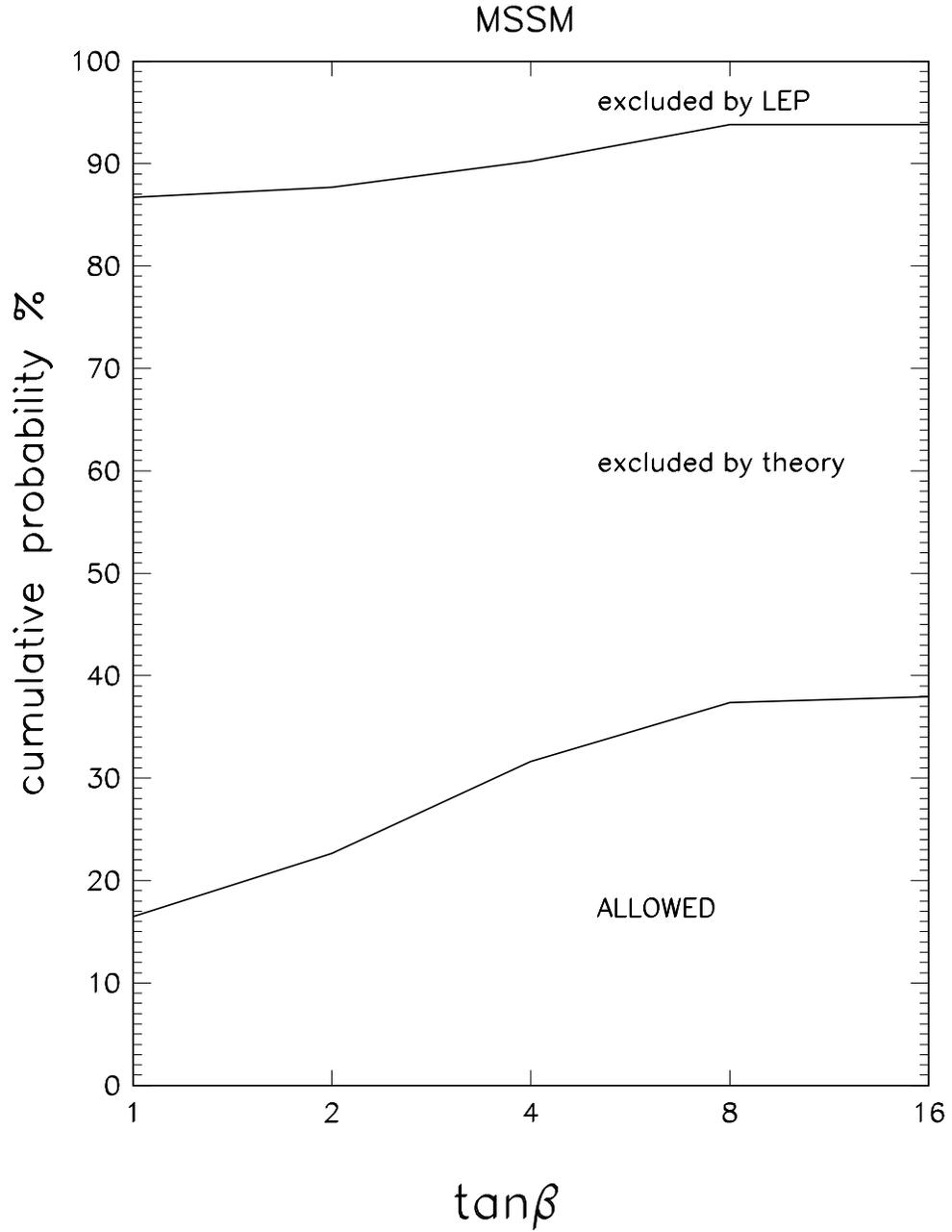

\caption{	Cumulative (integrated) probabilities in the various zones
		excluded or allowed in the Minimal Supersymmetric Standard 
		Model by one-sided bounds. No value of $\tan\beta$ can
		be excluded.}
\end{figure}
\begin{figure}
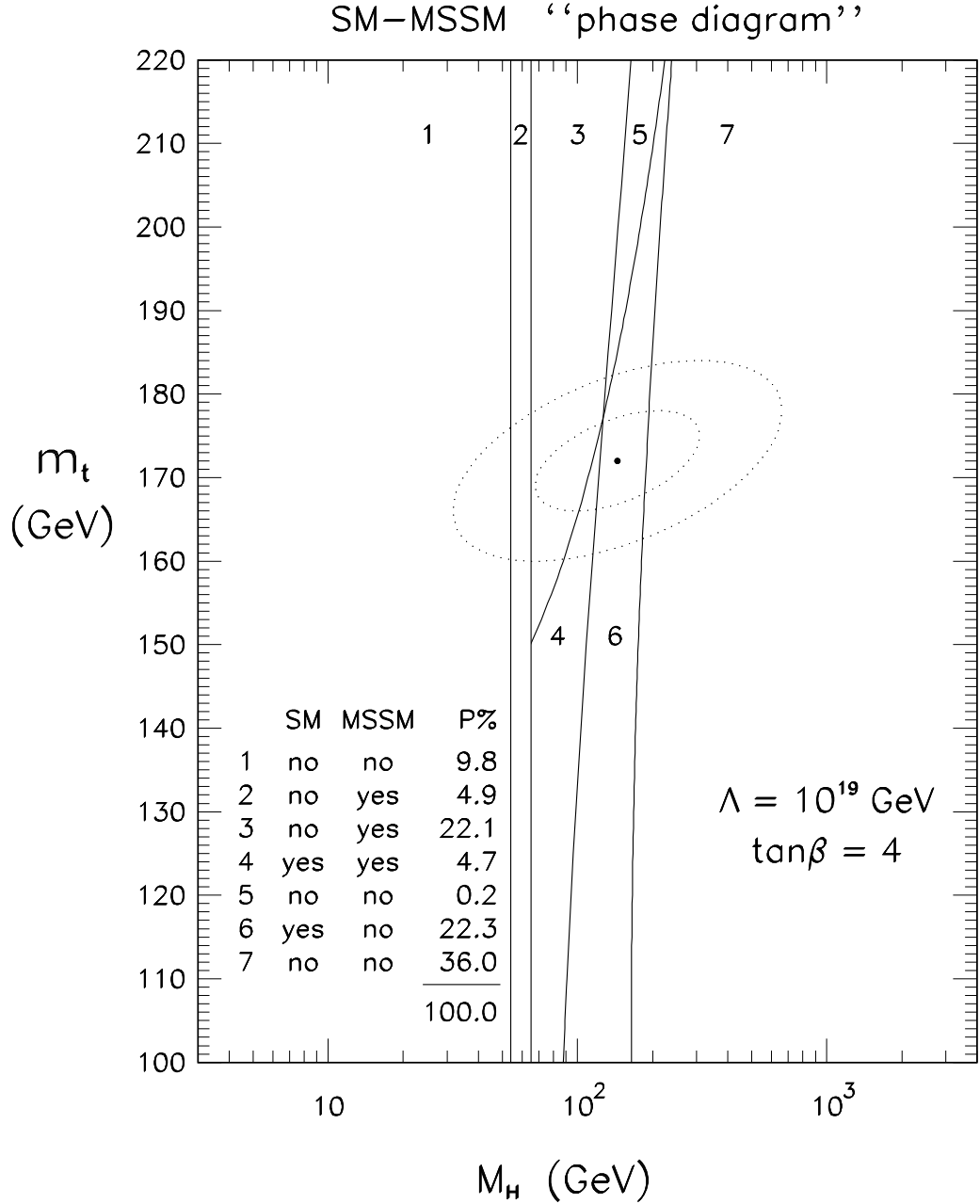

\caption{	Superposition of SM and MSSM bounds in the  $(M_H,\,m_t)$ 
		plane, for $\Lambda=10^{19}$ GeV and $\tan\beta=4$.
		The various zones 1--7 define regions that are (not) 
		compatible with a SM or a MSSM Higgs boson. The relative 
		likelihoods of these zones are estimated by ``weighting'' 
		them by the $(M_H,\,m_t)$ probability distribution, whose 
		1-$\sigma$ and 2-$\sigma$ contours are shown as dotted 
		ellipses. We also display the cumulative probability in 
		each zone.}
\end{figure}


\newcommand{\InsertFigure}[2]{\newpage\begin{center}\mbox{%
\epsfig{bbllx=1.4truecm,bblly=1.3truecm,bburx=19.5truecm,bbury=26.5truecm,%
height=21.truecm,figure=#1}}\end{center}\vspace*{-1.85truecm}%
\parbox[t]{\hsize}{\small\baselineskip=0.5truecm\hskip0.5truecm #2}}
\InsertFigure{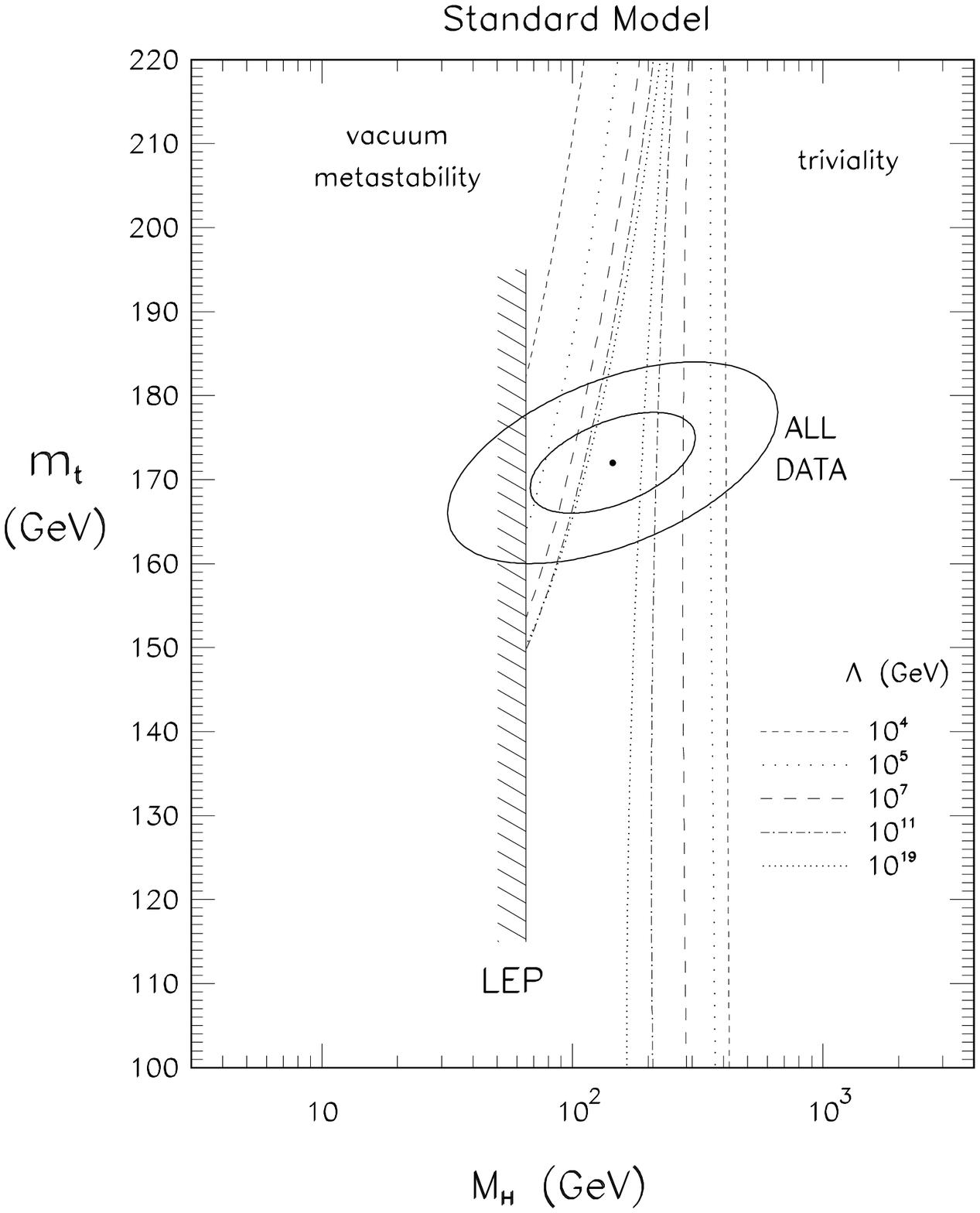}%
{FIG.~1. 	Indirect bounds on $(M_H,\,m_t)$ and one-sided experimental
		and theoretical limits in the Standard Model.  The solid 
		ellipses represent the 1-$\sigma$ and 2-$\sigma$ contours 
		from the best-fit  Gaussian distribution obtained by 
		analysing all electroweak precision data, including the 
		measurement of $m_t$ at CDF and D0. The hatched line is
		the LEP lower bound on $M_H$ \protect\cite{PDGR}. 
		The other curves represent the lower and upper limits on 
		$M_H$ from vacuum metastability  \protect\cite{Es95} and 
		triviality \protect\cite{Li86,Sh89} respectively, as 
		functions of the scale of new physics $\Lambda$.}
\InsertFigure{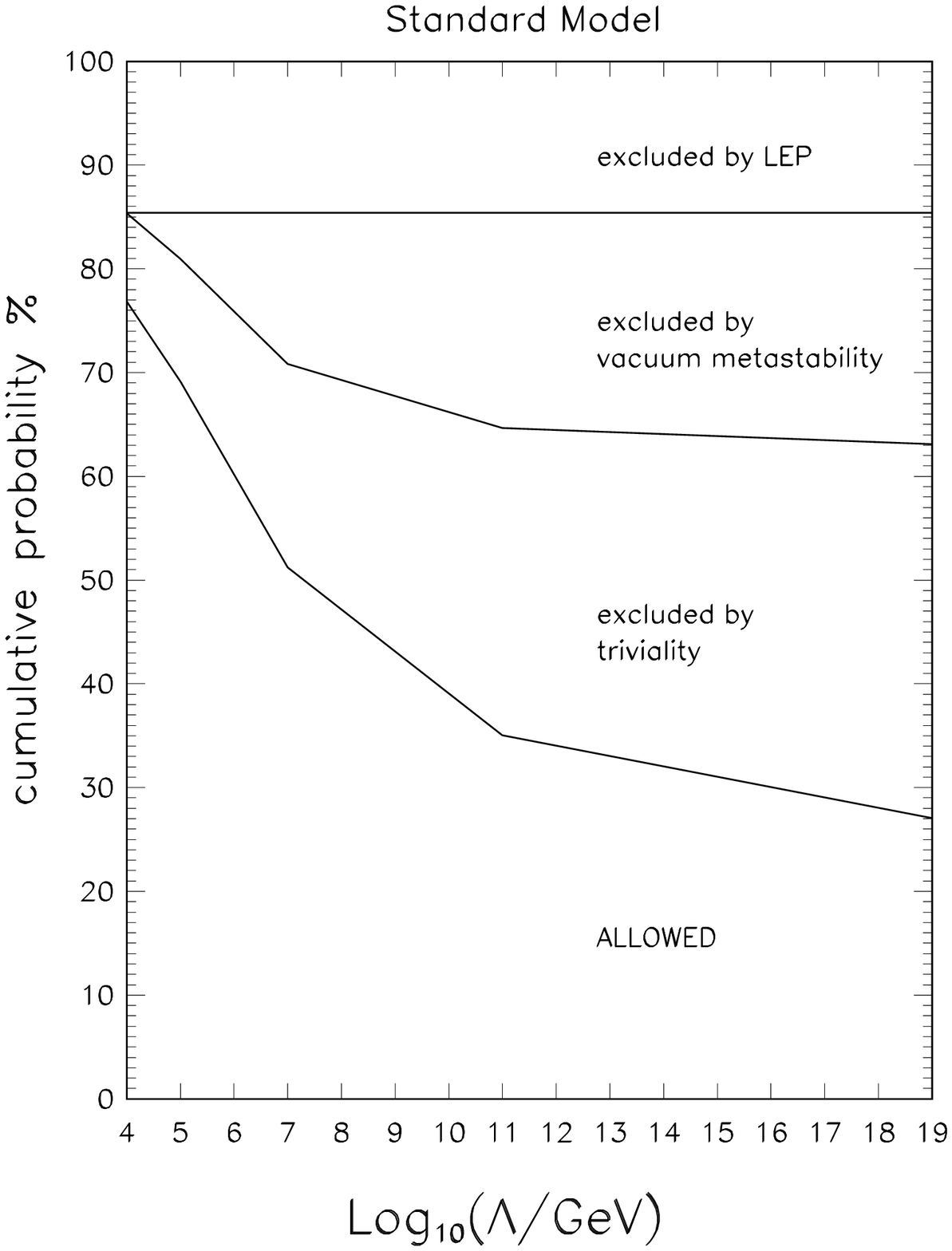}%
{FIG.~2. 	Cumulative (integrated) probabilities in the various zones
		excluded or allowed in the Standard Model by one-sided 
		bounds. No value of $\Lambda$ can be excluded.}
\InsertFigure{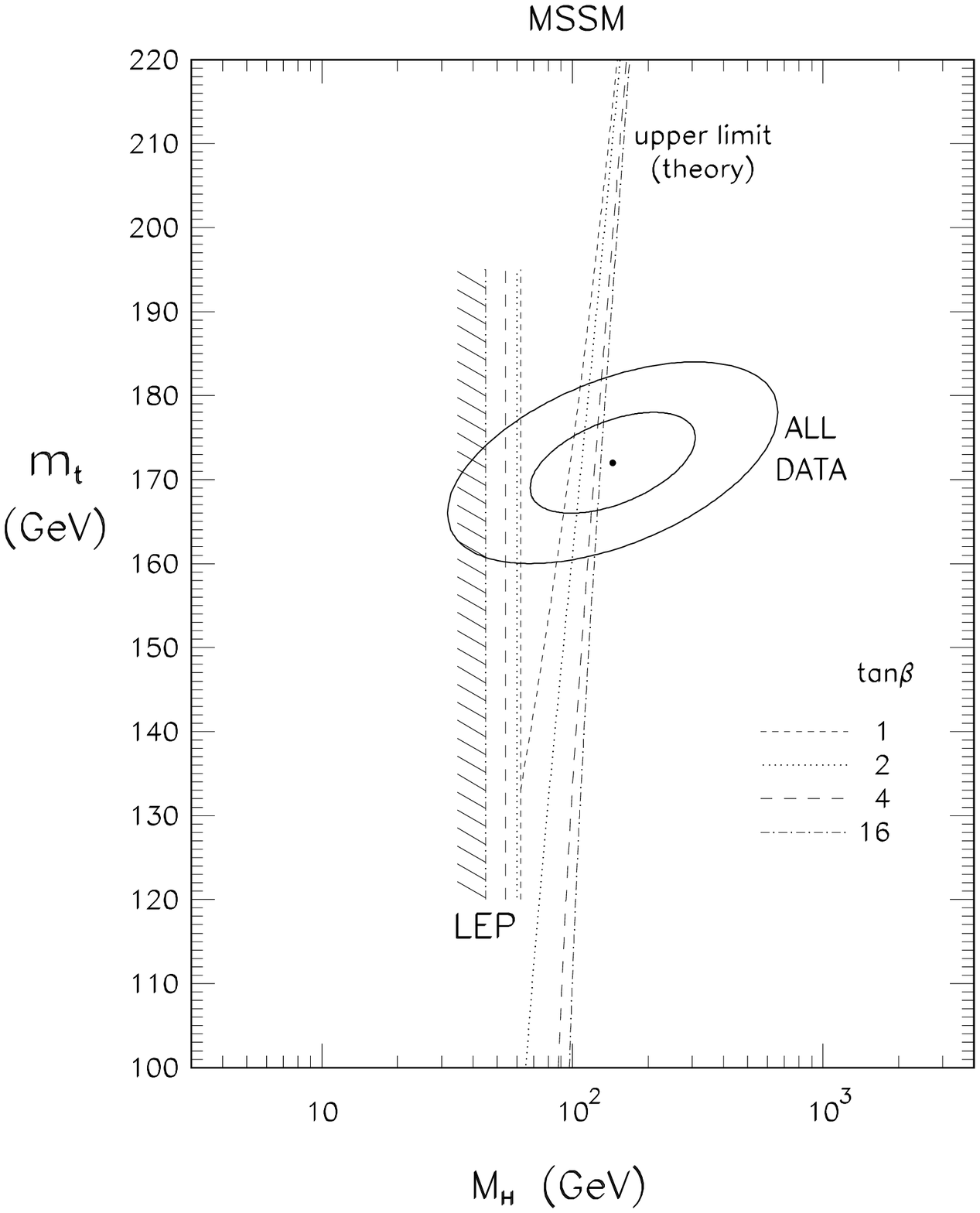}%
{FIG.~3. 	Indirect bounds on $(M_H,\,m_t)$ and one-sided experimental
		and theoretical limits in the Minimal Supersymmetric 
		Standard Model.  Apart from the Higgs sector, the MSSM 
		spectrum is assumed to be decoupled. Solid ellipses represent 
		the 1-$\sigma$ and 2-$\sigma$ contours as in Fig.~1. 
		The vertical lines are the LEP lower bounds on $M_H$ 
		\protect\cite{Gr95}, which depend slightly on $\tan\beta$. 
		The other curves represent the upper limits on $M_H$ in 
		the MSSM \protect\cite{MSSM}, as  a function of $\tan\beta$.}
\InsertFigure{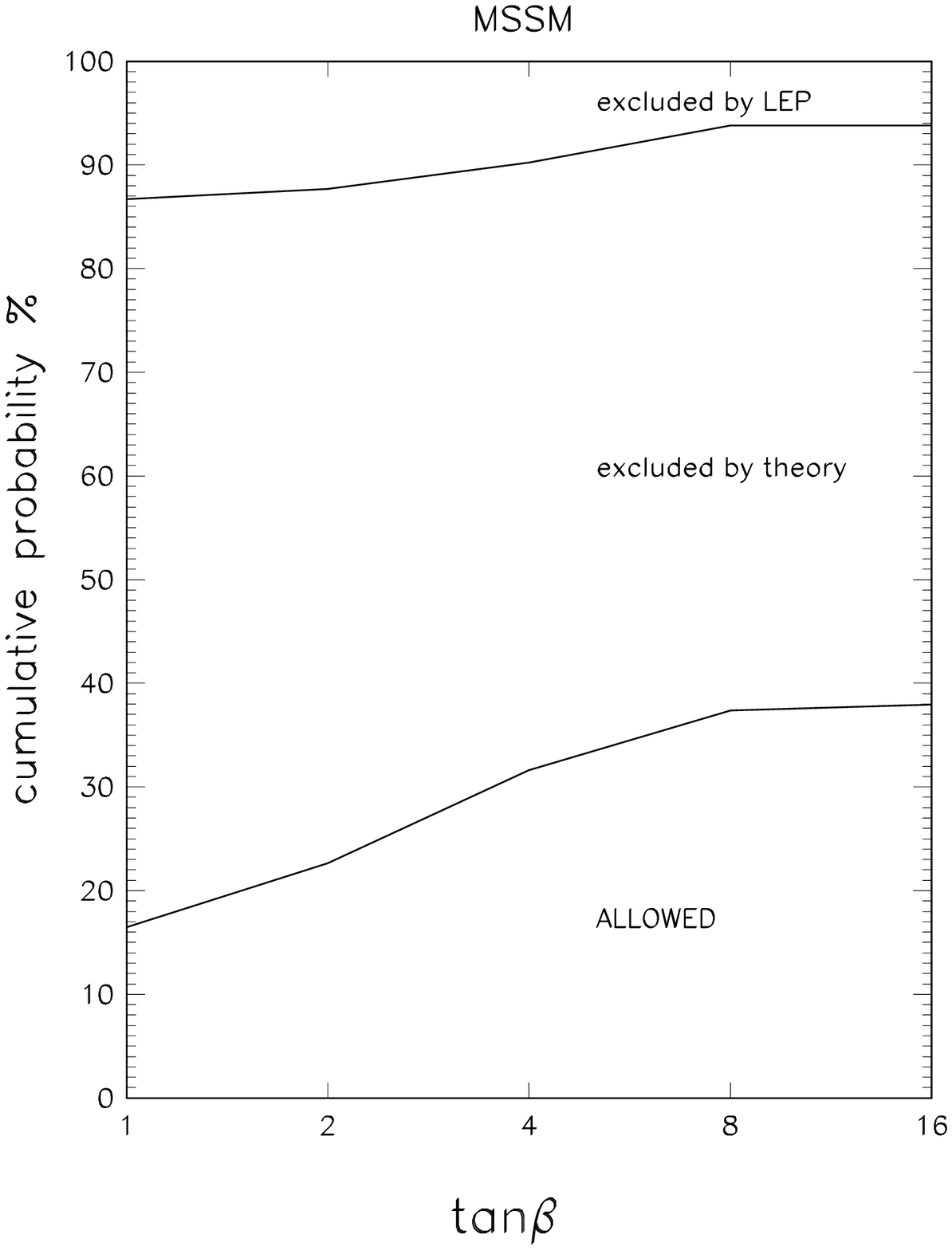}%
{FIG.~4. 	Cumulative (integrated) probabilities in the various zones
		excluded or allowed in the Minimal Supersymmetric Standard 
		Model by one-sided bounds. No value of $\tan\beta$ can
		be excluded.}
\InsertFigure{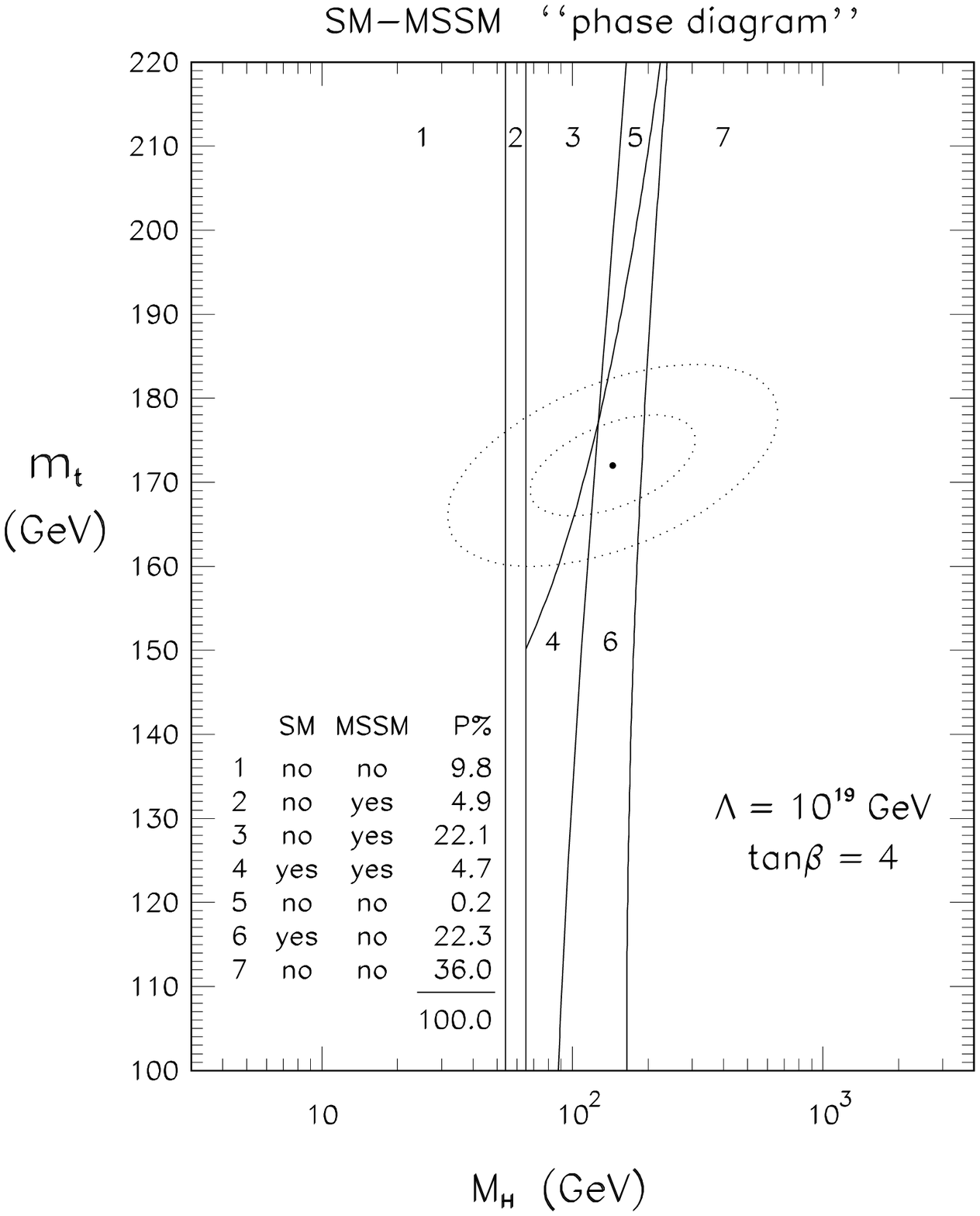}%
{FIG.~5. 	Superposition of SM and MSSM bounds in the  $(M_H,\,m_t)$ 
		plane, for $\Lambda=10^{19}$ GeV and $\tan\beta=4$.
		The various zones 1--7 define regions that are (not) 
		compatible with a SM or a MSSM Higgs boson. The relative 
		likelihoods of these zones are estimated by ``weighting'' 
		them by the $(M_H,\,m_t)$ probability distribution, whose 
		1-$\sigma$ and 2-$\sigma$ contours are shown as dotted 
		ellipses. We also display the cumulative probability in 
		each zone.}

\begin{thebibliography}{99}

\bibitem{El90}	J.~Ellis and G.~L.~Fogli,
		Phys.\ Lett.\ B {\bf 249} (1990) 543. 

\bibitem{El91}	J.~Ellis, G.~L.~Fogli, and E.~Lisi,
		Phys.\ Lett.\ B {\bf 274} (1992) 456.

\bibitem{El93}	J.~Ellis, G.~L.~Fogli, and E.~Lisi,
		Phys.\ Lett.\ B {\bf 318} (1993) 148.

\bibitem{Ag92}	F.~del Aguila, M.~Martinez, and M.~Quir{\'o}s,
		Nucl.\ Phys.\ B {\bf 381} (1991) 451.

\bibitem{No93}	V.~A.~Novikov, L.~B.~Okun, M.~I.~Vysotski, and V.~P.~Yurov,
		Phys.\ Lett.\ B {\bf 308} (1993) 123.

\bibitem{Ha94}	K.~Hagiwara, S.~Matsumoto, D.~Haidt, and C.~S.~Kim,
		Z.\ Phys.\ C {\bf 64} (1994) 559; {\it ibidem\/}
		C {\bf 68} (1995) 352(E).

\bibitem{Mo94}	G.~Montagna, O.~Nicrosini, G.~Passarino, and
		F.~Piccinini, Phys.\ Lett.\ B {\bf 335} (1994) 484.

\bibitem{El92}	J.~Ellis, G.~L.~Fogli, and E.~Lisi,
		Phys.\ Lett.\ B {\bf 286} (1992) 85.

\bibitem{El94}	J.~Ellis, G.~L.~Fogli, and E.~Lisi,
		Phys.\ Lett.\ B {\bf 333} (1994) 118. 

\bibitem{Li93}	J.~Ellis, G.~L.~Fogli, and E.~Lisi,
		Nucl.\ Phys.\ B {\bf 393} (1993) 3.

\bibitem{Ch96}	P.~H.~Chankowski and S.~Pokorski,
		Phys.\ Lett.\ B {\bf 366} (1996) 188. 

\bibitem{Ho96}	W.~de Boer, A.~Dabelstein, W.~Hollik,
		and W.~M{\"o}sle,
		Karlsruhe University Report No.\ KA-TP-18-96,
		hep-ph/9607286. 

\bibitem{Ctop}	CDF Collaboration, F.~Abe {\it et al.\/},
		Phys.\ Rev.\ Lett.\ {\bf 74} (1995) 2676.

\bibitem{Dtop}	D0 Collaboration, S.~Abachi {\it et al.\/},
		Phys.\ Rev.\ Lett.\ {\bf 74} (1995) 2632.

\bibitem{EWWG}	A.~Blondel, Plenary talk at the International Conference
                on High Energy Physics, Warsaw, 1996, reporting the analysis 
		of the LEP Electroweak Working Group and the SLD Heavy
		Flavor Group, CERN Report No.\ LEPEWWG/96-02, available at
		the URL: http://www.cern.ch/LEPEWWG .

\bibitem{El96}	J.~Ellis, G.~L.~Fogli, and E.~Lisi,
		Z.\ Phys.\ C {\bf 69} (1996) 627. 

\bibitem{Ch95}	P.~H.~Chankowski and S.~Pokorski,
		Phys.\ Lett.\ B {\bf 356} (1995) 307; see also the update,
		hep-ph/9509207.

\bibitem{Ma95}	S.~Matsumoto, Mod.\ Phys.\ Lett.\ A {\bf 10} (1995) 2553.

\bibitem{PDGR}	Particle Data Group, R.~M.~Barnett {\it et al.\/},
		Phys.\ Rev.\ D {\bf 54} (1996) 1.

\bibitem{Gr95}	J.F. Grivaz,
		Rapporteur talk given at {\it HEP~'95},
		International	 Europhysics Conference on High Energy 
		Physics, Brussels, Belgium, 1995.

\bibitem{Es95}	J.~R.~Espinosa and M.~Quir{\'o}s,
		Phys.\ Lett.\ B {\bf 353} (1995) 257. 

\bibitem{Li86}	M.~Lindner, Z.\ Phys.\ C {\bf 31} (1986) 295.

\bibitem{Sh89}	M.~Sher, Phys.\ Rept.\ {\bf 179} (1989) 273. 

\bibitem{MSSM}	Y.~Okada, M.~Yamaguchi and T.~Yanagida,
                Progr.\ Theor.\ Phys.\ {\bf 85} (1991) 1;
                J.~Ellis, G.~Ridolfi and F.~Zwirner,
                Phys.\ Lett.\ B {\bf 257} (1991) 83 and B {\bf 262} 
                (1991) 477;
                H.~E.~Haber and R.~Hempfling,
                Phys.\ Rev.\ Lett.\ {\bf 66} (1991) 1815. 

\bibitem{Di95}	M.~A.~Diaz, T.~A.\ ter Veldhuis, and T.~J.~Weiler,
		Phys.\ Rev.\ Lett.\ {\bf 74} (1995) 2876; see also
		the updated version, Vanderbilt University Report
		No.\ VAND-TH-94-14-UPD, hep-ph/9512229; P.~Q.~Hung
		and M.~Sher, Phys.\ Lett.\ B {\bf 374} (1996) 138.

\bibitem{Es96}	J.~R.~Espinosa, DESY Report No.\ 96-107, hep-ph/9606316,
		Lecture at {\it ITEPWS~'96\/},
		XXIV ITEP Winter School of Physics, to appear in 
		the Proceedings.

\bibitem{Ka95}	K.~Kang and S.~K.~Kang, talk at ``Beyond the Standard
		Model IV,'' Granlibakken, Lake Tahoe, California (1994),
		hep-ph/9503478.

\end{thebibliography}
\end{document}